\documentclass{article}
\pdfoutput=1

\PassOptionsToPackage{numbers}{natbib}

\usepackage[preprint]{nips_2018}

\usepackage[utf8]{inputenc} 
\usepackage[T1]{fontenc}    
\usepackage{hyperref}       
\usepackage{graphicx}
\usepackage{multicol}
\usepackage{framed}
\usepackage{subcaption}
\usepackage{url}            
\usepackage{booktabs}       
\usepackage{amsfonts}       
\usepackage{nicefrac}       
\usepackage{microtype}      
\usepackage{amsmath}

\title{VDMS: Efficient Big-\textit{Visual}-Data Access for \\
       Machine Learning Workloads}

\author{
  Luis Remis \thanks{Corresponding author - \texttt{luis.remis@intel.com}} \\
  Intel Labs\\
  \And
  Vishakha Gupta-Cledat \\
  Intel Labs\\
  \And
  Christina R. Strong \\
  Intel Labs\\
  \And
  Ragaad Altarawneh \\
  Intel Labs\\
}

\begin{document}

\maketitle

\begin{abstract}
We introduce the Visual Data Management System (VDMS),
which enables faster access to big-{\em visual}-data
and adds support to visual analytics.
This is achieved by searching for relevant visual data via metadata stored as
a graph, and enabling faster access to visual data through new
machine-friendly storage formats. VDMS differs from existing large scale photo
serving, video streaming, and textual big-data management systems due to
its primary focus on supporting machine learning and
data analytics pipelines that use visual data
(images, videos, and feature vectors), treating these as first class entities.
We describe how to use VDMS via its user friendly interface and how
it enables rich and efficient vision analytics through a machine learning
pipeline for processing medical images. We show the improved
performance of 2x in complex queries over a comparable set-up.
\end{abstract}

\section{Introduction}
\label{intro}

Visual computing workloads performing analytics on
video or image data, either off-line or streaming,
have become prolific across a wide range of application domains.
This is in part due to the growing ability of machine learning techniques to
extract information out of the visual data which can subsequently be used
for informed decision making.
The insights this information can provide depend on the
application: a retail vendor might be interested in the amount of time
shoppers spend in front of a specific product, while a medical expert might
want to see the effect of a specific treatment on the size of a tumor.

Despite this rich and varied usage environment, there has been very little
research on the management of visual data.
Most of the current storage solutions are
an ad-hoc collection of tools combined with custom scripts to tie them
together, unique not only to a specific discipline but often to individual
researchers. For example, consider an ML developer constructing a pipeline
for extracting brain tumor information from existing brain images in a
classic medical imaging use case. This requires assigning consistent
identifiers for the scans and adding their metadata in
some form of relational or key-value database. If the queries require
search over some patient information, then patients have to be associated
with their brain scans. Finally, if the ML pipeline needs images that
are of a size different than the stored ones, there is additional compute
diverted towards preprocessing after the potentially larger images are
fetched. All these steps require investigation of different software
solutions that provide various functionalities that can then be stitched
together with a script for this specific use case.
Moreover, if the pipeline identifies
new metadata to be added for the tumor images, most databases make it
hard to evolve the schema on the fly.
Not only do these ad hoc solutions make replicating experiments
difficult, they do not scale well to real-world applications.
Addressing the storage and retrieval of visual data necessitates a complete
overhaul of the storage architecture,
preferably using emerging breakthroughs in
heterogeneous memory and storage for efficiency.

We present the Visual Data Management System (VDMS)\cite{darkside},
an Open Source project designed to enable efficient access of visual data.
Since visual data often contains
rich metadata (such as objects, locations, and time), VDMS stores this
information in a high performance graph database. Using this metadata, VDMS
can quickly identify which data is relevant to a given query.
Additionally, VDMS uses a custom library to store and retrieve visual data,
which provides an interface for machine friendly formats as well as
traditional formats. These new formats are designed to support applications
that are often interested in specific areas of images or videos,
particularly when the individual object is large.

While there are a number of big-data frameworks~\cite{spark, hadoop}, systems
that can be used to store metadata~\cite{memsql, vertica}, and systems that
manipulate a specific category of visual data~\cite{scidb, rasdaman}, VDMS can
be distinguished from them on the following aspects:

\begin{itemize}
\item {\em Design for analytics and machine learning}: by targeting
visual data for use cases that require manipulation
of visual information and associated metadata,
\item {\em Ease-of-use}: By defining a common API that allows applications to
combine their complex metadata searches with operations on resulting visual
data, and together with full support for feature vectors, VDMS goes beyond the
traditional SQL or OpenCV level interfaces that do one or the other. Given our
focus on enabling machine learning, we also provide a client API in Python.
\item {\em Performance}: We show how a unified system such as VDMS can
outperform an ad-hoc system constructed with well-known discrete components.
Because of the capabilities we have built into VDMS, it handles complex
queries significantly better than the ad-hoc system without compromising the
performance of simple queries.
\end{itemize}

\section{Design and Implementation}
\label{arch}

VDMS implements a client-server design.
The VDMS Server handles client requests concurrently and coordinates
request execution across the metadata and data components in order to return a
unified response. The metadata component is the Persistent Memory Graph
Database (PMGD). The data component is our custom library, the
Visual Compute Library (VCL). The VCL enables machine friendly enhancements to
visual data.
VDMS and its components are fully available as open source tools
\footnote{https://github.com/IntelLabs/\{vdms, pmgd, vcl\}}.

\textbf{Persistent Memory Graph Database:}
Recent developments in persistent memory technologies
like 3D XPoint~\cite{IntelXPoint15}
promise storage elements  providing  nearly  the  speed  of  DRAM  and  the
durability of block-oriented storage. To provide an efficient storage
solution addressing the increasing popularity of connected data and
applications that benefit from graph like processing, we have designed
and implemented an in-persistent-memory graph database, PMGD, optimized
to run on a platform equipped with persistent memory.
PMGD provides a property graph model of data storage with the traditional
atomicity, consistency, isolation, and durability properties expected from
databases. The graph model makes it very suitable for the data model and
access patterns shown by visual metadata.
With its natural ability to extend the schema very
easily (due to the use of a property graph model),
we can support new developments in machine learning that can lead to
enhancements to existing metadata over time.
The specific details on how the data structures are optimized for persistent
memory, and the performance comparison with other graph databases is on-going
work that will be publish separately.

\textbf{Visual Compute Library:} The VCL was designed to provide an interface
through which users can interact with visual data. For traditional formats,
the interface is an abstraction layer over OpenCV. However, it also provides a
way to use novel formats that are better suited for visual analytics: a novel,
array-based lossless image format. This format is built on the array data
manager TileDB~\cite{TileDB} and is well suited for images that are used in
visual analytics.
The VCL currently provides limited support for videos but we are enhancing
its capabilities as part of our future work.
Feature vector support is provided through an implementation based
on high-dimensional sparse arrays, also using TileDB, which contains both
storage and search functionality over feature vectors.
In addition, the VCL provides a wrapper
for another high-dimensional index implementation,
Facebook's Faiss~\cite{faiss}.

\begin{figure}[ht!]
\begin{subfigure}{.5\textwidth}
  \centering
  \includegraphics[width=.92\linewidth]{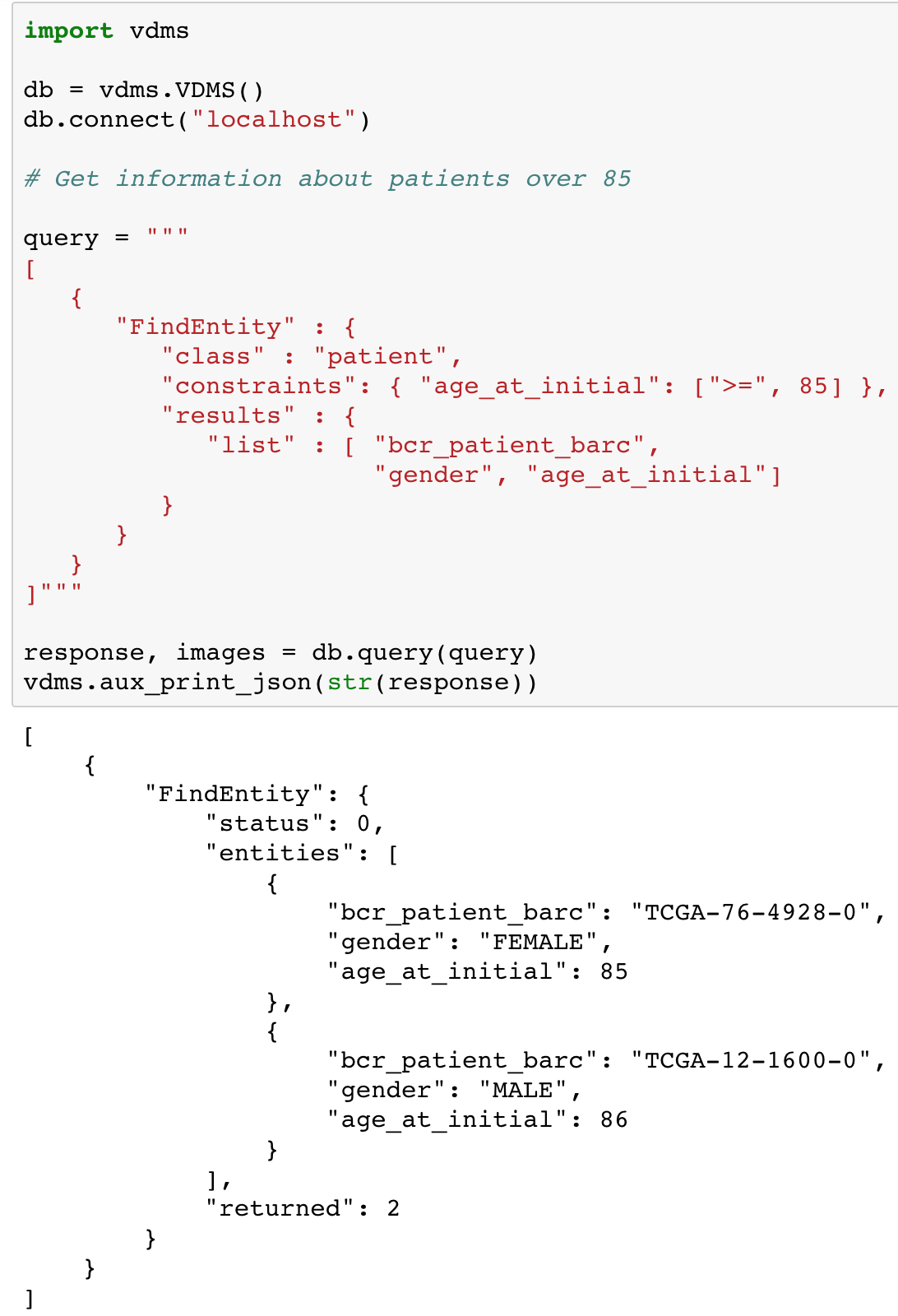}
  \caption{Simple Metadata Query.}
  \label{fig:sfig1}
\end{subfigure}%
\begin{subfigure}{.52\textwidth}
  \centering
  \includegraphics[width=.94\linewidth]{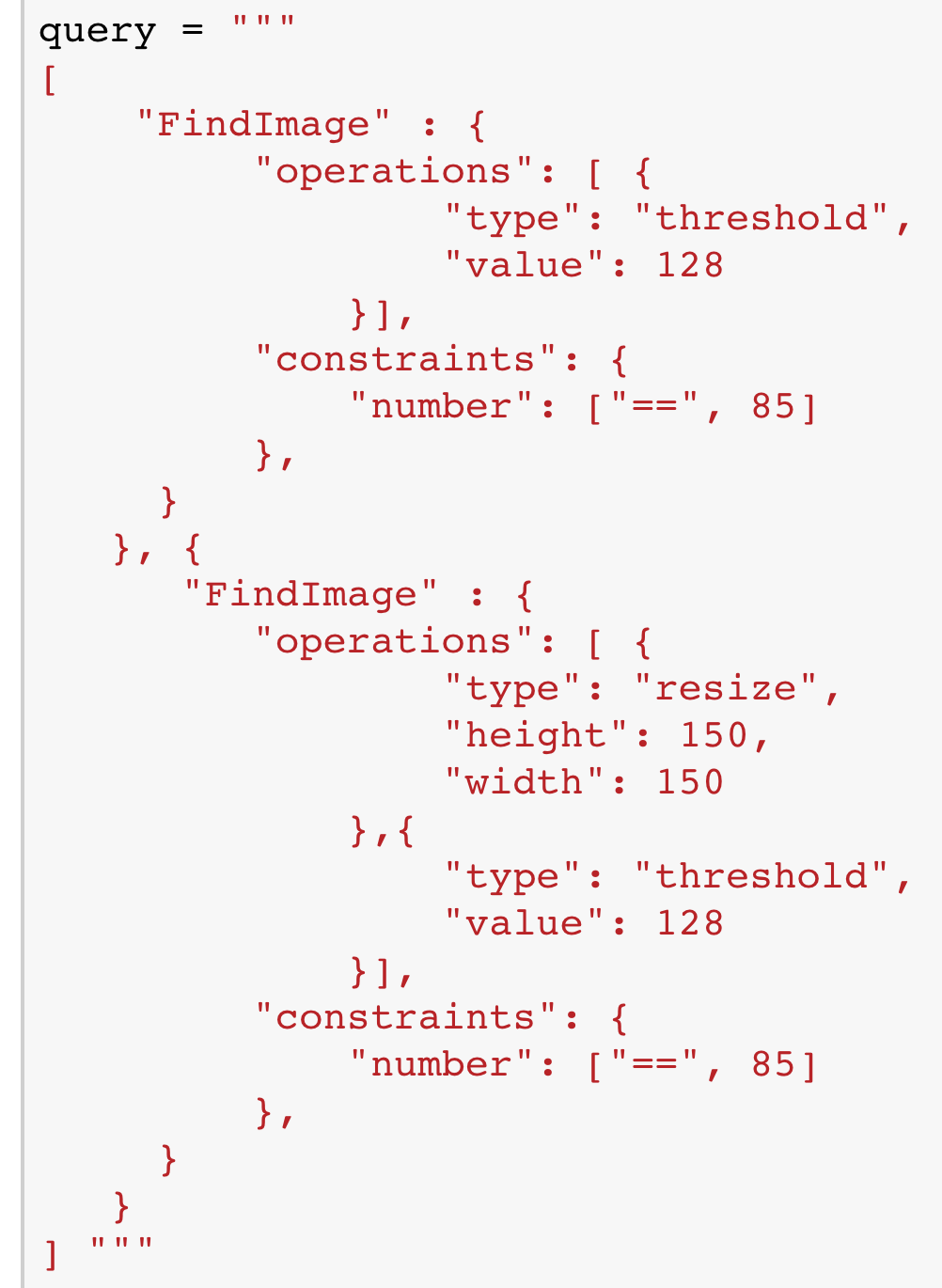}
  \caption{Query involving visual transformations.}
  \label{fig:sfig2}
\end{subfigure}
\begin{subfigure}{\textwidth}
  \centering
  \includegraphics[width=.9\linewidth]{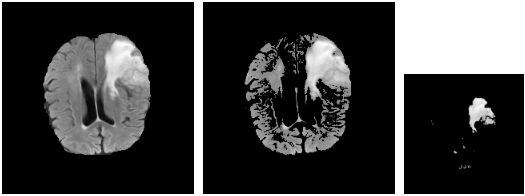}
  \caption{Original image (left) and the result of query in (b)}
  \label{fig:sfig3}
\end{subfigure}
\caption{Snapshot of a Python notebook with sample queries}
\label{fig:query}
\end{figure}

\textbf{Request Server:}
Developers and users of machine learning frameworks and data science
applications favor simpler interfaces to access and process data and cannot
be expected to deal with two different ways of interacting with information (
metadata and visual data) instead of focusing on the algorithmic parts of their
pipelines.
VDMS takes care of coordinating client requests across the metadata and the
data as well as efficiently manages multiple clients through its Request
Server component, by implementing a JSON-based API.
It decomposes the command into
metadata and data requests, invokes the relevant calls behind the scene,
and returns a coherent response to the user after applying any additional
operations (explained in the next section).


\section{VDMS API}
\label{arch}
One of our goals with VDMS was to define an API that is easy to use.
Our API explicitly predefines certain
visual primitives associated with images, videos, and feature vectors. In
addition, while we use a graph database to store our metadata, the API is not
graph specific.
We have paid particular attention to hide the complexities of our internal
implementation and up-level the API to a JSON-based
API\footnote{https://github.com/IntelLabs/vdms/wiki/API-Description},
which is very popular across various application domains.
We understand that by defining a new JSON-based API we had to trade-off
expressiveness (compared to SPARQL or Gremlim) for the ability to design
native support for visual data, but we believe that, as development of VDMS
evolves, we will be able to achieve similar levels of expressiveness compared
to more mature query languages.
We have developed Python and C++ client
library to provide a simple query function that accepts a JSON string with
commands and an array or vector of blobs.

\begin{figure}[ht!]
\centering
\includegraphics[width=0.9\columnwidth]{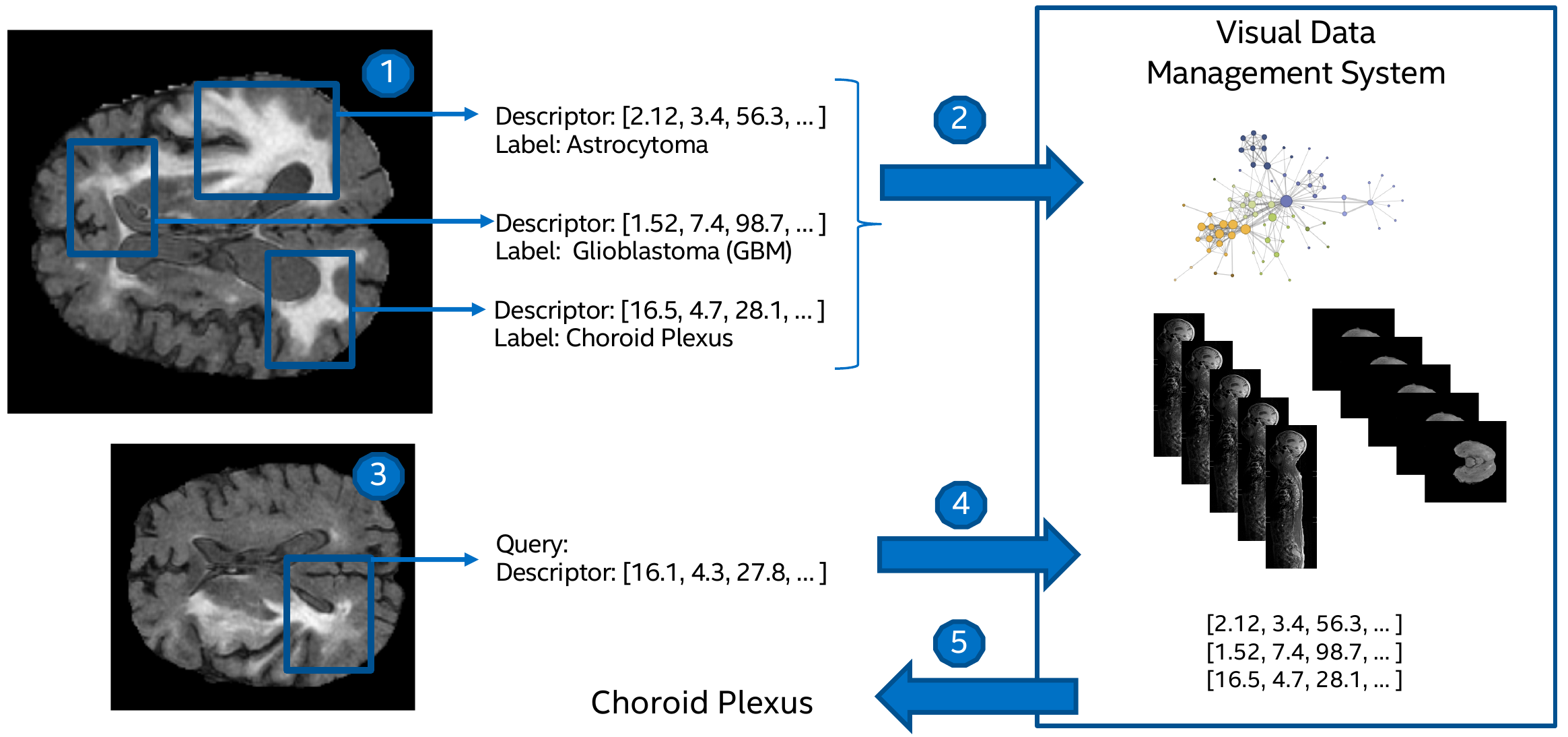}
\caption{Feature Vectors natively supported in VDMS}
\label{fig:features}
\end{figure}

We use a medical imaging dataset as a driving example, and use it
here to showcase our API. More details about this use case are discussed in
Section~\ref{demo}.
Figure~\ref{fig:query} shows a snapshot from a Jupyter Notebook with two
simple queries.
Figure~\ref{fig:sfig1} shows a simple metadata query
where information about properties of patients matching some specific
characteristics (admitted with 85 years of age or more)
is used to constrain the search.
The returned JSON structure shows 2 patients found in the database.
Note that in this examples the queries are expressed as plain strings to
help the reader understand its structure. In real applications, the queries
are structured using the preferred method or library
for handling JSON structures (e.g. Python dictionaries or jsoncpp in C++).

Figure~\ref{fig:sfig2} shows a query involving visual transformations.
In this case, the query is looking for an image with a certain \textit{id}
value, and expects  VDMS to return the image twice: once with a thresholding
transformation (e.g., zero out all pixels less than a threshold), and the
second time after applying a threshold and then resizing it.
\ref{fig:sfig3} shows the image originally inserted in VDMS (left),
along with the returned images corresponding to the query in \ref{fig:sfig2}.
Adding new operations to VDMS is easy, as our software architecture
encapsulates all operations in VCL, and any new operation can use
or wrap around OpenCV, which is used internally.

Another key differentiating factor of VDMS is that it allows the creation of
indexes for high-dimensional feature vectors and the insertion of
these feature vectors associated with entities or visual objects.
Feature vectors are intermediate results of various machine
learning or computer vision algorithms when run on visual data.
These vectors can be labeled and classified to build search indexes,
and there are many in-memory libraries that are designed for
this task~\cite{flann, faiss}.
Using the VDMS API, users can manage feature vector indexes,
query previously inserted elements (images),
run a k-nearest neighbor search (knn) and express relationships
between existing images or feature vectors and
the newly inserted feature vectors.
By natively supporting feature vectors and knn,
VDMS allows out-of-the-box classification
functionalities for many applications.
Figure~\ref{fig:features} shows an example of how this functionality
can be integrated in a medical imaging analytics pipeline:
(1) A feature vector is extracted from an bounding
box in a brain image and labeled;
(2) the feature vector is inserted together
with all the associated metadata (type of cancer, for example)
and a link to the original image.
(3) A new feature vector is extracted from a new image,
(4) a query to VDMS is issued to classify that feature vector based on the
indexed features in VDMS, and
(5) VDMS can respond to the query with the label
associated to that feature vector.
Note that, even if the example is based on the medical images used for the
performance evaluation, this methodology is applicable in many other contexts
and use cases, such as face detection and matching.
More details about the JSON API for this functionality can be found on
our Github wiki page.

\section{Medical Imaging Use Case}
\label{demo}

Standards for medical images, such as DICOM and NIfTI, were developed to assist
in transmitting medical images along with their associated metadata. This
metadata, consisting of patient
information and often including treatments, can be
used together with the images themselves
in machine learning applications to gather insights.
This makes medical imaging an
excellent use case to demonstrate the metadata query and data pre-processing
capabilities offered by VDMS. It also helps us explain how queries can be made
increasingly complex to give more detailed information, finding the most
relevant subset of patients for focused analyses.

\begin{figure}[htb]
\centering
\includegraphics[width=0.9\columnwidth]{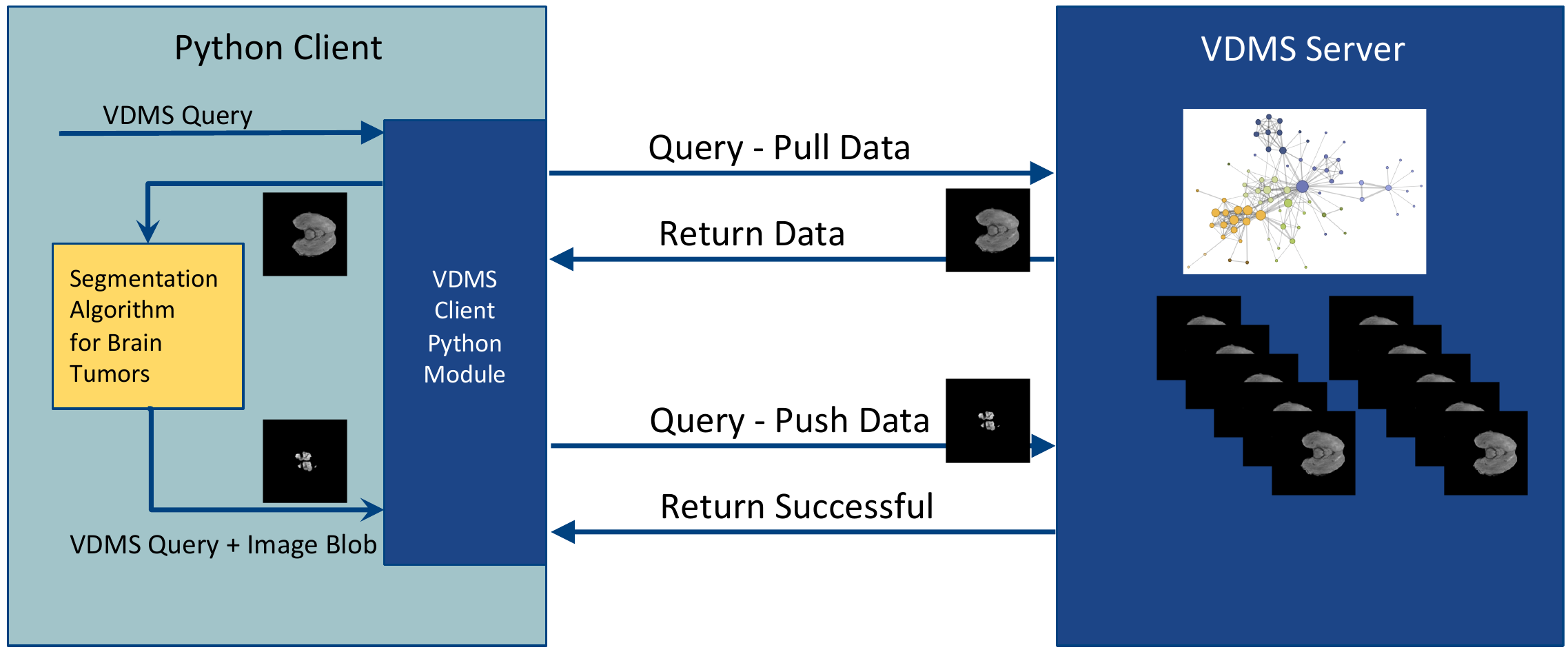}
\caption{Segmentation pipeline to find brain tumors in existing patient brain scans}
\label{fig:segpipeline}
\end{figure}

We implemented a pipeline for a medical imaging use case
that processes brain scans. This pipeline feeds brain images to a convolution
neural network (using U-Net) that runs a segmentation,
the results of which are pushed
back to storage for future use, as shown in Figure \ref{fig:segpipeline}.
We use The Cancer Image Archive dataset~\cite{tcia}
for both metadata and image information (DICOM files).

We present our findings after performing three queries based on both
metadata and visual data (brain scans) that are basic building blocks for
running the pipeline in Figure \ref{fig:segpipeline}. All queries return
a resized (downsampled) version of the image that
matched the requirement of the input layer of the CNN.

The three queries are:
\begin{enumerate}
\item {\bf Query 1}: retrieve a single image from a brain scan, search by its unique name,
and apply pre-processing operations: Simple case where a single
command on the API is used.

\item {\bf Query 2:} retrieve a complete brain scan (155 images)
from a particular patient,
and apply pre-processing operations: Involves performing a neighbor search
on the metadata graph, plus accessing multiple images.

\item {\bf Query 3:} retrieve all brain scans corresponding to people over 75
who had a chemotherapy using the drug \textit{Temodar}:
Involves performing a more complex metadata
graph traversal, plus accessing images corresponding to multiple brain scans.
\end{enumerate}

\subsection{Performance Evaluation}

As a baseline, we implemented a similar set-up, in terms of functionalities,
using off-the-shelf, popular, components: MemSQL Server for metadata
storage, Apache HTTP Web server for image storage
(comparable in functionalities/interfaces to a cloud blob store),
and OpenCV for pre-processing operations
(pre-processing is a necessary step before inputing to the CNN).
We chose this approach because there are no
other systems that integrate all the needed functionalities to provide
visual data access together with metadata for this segmentation pipeline.
We focus the performance evaluation on the data access, given that
it is increasingly becoming a problem on the overall execution time
when dealing with analytics pipelines \cite{darkside}.

We use the following configuration (both for VDMS and the baseline):
Intel\textsuperscript{\textregistered}
Xeon\textsuperscript{\textregistered}
Gold 6140 CPU @ 2.30GHz CPU server runs in a Linux environment,
running Ubuntu 16.04, Python 2.7, and gcc 5.4. The same configuration
was used for both server and client machines, connected over a 1Gbps link.

\begin{figure}[htb]
\centering
\includegraphics[width=1\columnwidth]{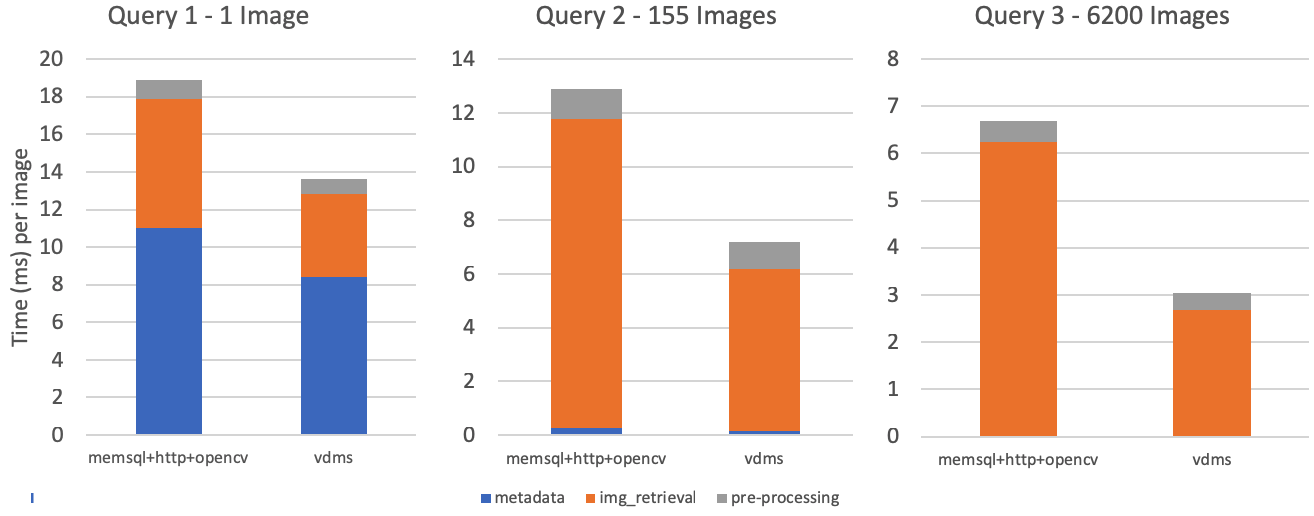}
\caption{Performance evaluated using the medical imaging queries}
\label{fig:perf}
\end{figure}

Figure~\ref{fig:perf} shows a sample from the performance evaluation we have
done comparing VDMS to the ad-hoc baseline, using the medical imaging
queries described above.
\textit{metadata} represents the time spent in querying metadata only,
\textit{img\_retrieval} represents the time spent on
image read from disk plus sending over the network, and
\textit{pre-processing} represents time spent when resizing the image.
In the case of VDMS, pre-processing happens in the server side.
We demonstrate how VDMS is able to perform very well
for complex queries without hurting the performance of simple queries.
VDMS significantly benefits from
co-locating pre-processing operations on images before data is
transferred to the user-application, something that is enabled by its
unified API.

For instance, in the case of Query 3, a complex query
that involves graph traversal and transfer of a large number of images,
VDMS is more than 2x faster than the baseline.
It can be seen that a large part of the performance
improvement comes from the fact that
certain operations, like resize to downsample an image, reduce the overall
amount of data that has to be transmitted between the client and the server.
That effect becomes more visible as more images are retrieved.
When more images are retrieved, the metadata performance have less effect
in the overall execution time, as it was expected.

Besides its performance, it is important to note that setting up VDMS
to work directly with the segmentation pipeline is significantly simpler than
having to deal with the three systems used for the baseline all together, and
we believe that together with the performance results showed, greatly
justify its use and adoption in real and critical scenarios.

\section{Conclusion}

We introduced the Visual Data Management System,
designed to enable efficient access of visual data.
We presented our rich, JSON-based API, designed to simplify visual data access
for data scientists and analytics pipelines.
We compared the performance of our system to a
baseline of a combination of widely available and used systems.
Our findings showed that VDMS efficiently deals with complex queries,
providing a performance improvement
of up to 2x in the examined medical data search use-case.
Furthermore, VDMS requires significantly fewer
lines of code to execute complex queries in complex visual pipelines.
We intend to continue with the evaluation of
VDMS performance different use cases, and to identify more
opportunities to optimize and simplify the access of visual data
for machine learning and analytics applications.

\begin{framed}
In Memoriam to Scott Hahn, who guided us through this journey.

He will always be in our hearts and memories.
\end{framed}


\bibliographystyle{abbrv}
\bibliography{vdms}

\scriptsize
\begin{framed}

2018 Intel Corporation.

Intel and Intel Core are trademarks of Intel Corporation in
the U.S. and/or other countries. Other names and brands may be claimed as the property of others.

Performance results are based on testing as of 10/01/2018 and may not reflect
all publicly available security updates.
See configuration disclosure for details.
No product can be absolutely secure.

Software and workloads used in performance tests may have been optimized for
performance only on Intel microprocessors. Performance tests, such as SYSmark
and MobileMark, are measured using specific computer systems, components,
software, operations and functions.
Any change to any of those factors may cause
the results to vary. You should consult other information and performance tests
to assist you in fully evaluating your contemplated purchases, including the
performance of that product when combined with other products. For more
information go to http://www.intel.com/performance.

Intel's compilers may or may not optimize to the same degree for non-Intel
microprocessors for optimizations that are not unique to Intel
microprocessors.
These optimizations include SSE2, SSE3, and SSSE3 instruction
sets and other optimizations. Intel does not guarantee the availability,
functionality, or effectiveness of any optimization on microprocessors not
manufactured by Intel. Microprocessor-dependent optimizations in this product
are intended for use with Intel microprocessors. Certain optimizations not
specific to Intel microarchitecture are reserved for Intel microprocessors.
Please refer to the applicable product User and Reference Guides for more
information regarding the specific instruction sets covered by this notice.
\end{framed}

\end{document}